\documentclass[12pt,a4paper]{article} 
\usepackage[dvips]{graphicx} 
\newcommand{\lsim}{\mathrel{\lower4pt\hbox{$\sim$}}
                   \hskip-12.5pt\raise1.6pt\hbox{$<$}\;}
\newcommand{\gsim}{\mathrel{\lower4pt\hbox{$\sim$}}
                   \hskip-12.5pt\raise1.6pt\hbox{$>$}\;}
\begin{document} 
\begin{titlepage} 
\begin{flushright} 
LNF--01/011(P)\\ 
ULB--TH/01--01\\ 
hep-ph/0105211\\ 
May 2001 
\end{flushright} 
\vspace*{1.6cm} 
 
\begin{center} 
{\Large\bf Insights on neutrino lensing}\\ 
\vspace*{0.8cm} 
 
R.~Escribano$^a$, 
J.-M.~Fr\`ere$^{b,}$\footnote{Directeur de recherches du FNRS.}, 
D.~Monderen$^b$ and V.~Van Elewyck$^b$\\ 
\vspace*{0.2cm} 
 
{\footnotesize\it  
$^a$INFN-Laboratori Nazionali di Frascati, 
P.O.~Box 13, I-00044 Frascati, Italy\\ 
$^b$Service de Physique Th\'eorique, Universit\'e Libre de 
Bruxelles, CP 225, B-1050 Bruxelles, Belgium}\ 
\end{center} 
\vspace*{1.0cm} 
 
\begin{abstract} 
We discuss the gravitational lensing of neutrinos by astrophysical objects.  
Unlike photons, neutrinos can cross a stellar core;  
as a result, the lens quality improves. 
We also estimate the depletion of the neutrino flux after crossing 
a massive object and the signal amplification expected. 
While Uranians alone would benefit from this effect in the Sun,  
similar effects could be considered for binary systems. 
\end{abstract} 
\end{titlepage} 
 
\section{Introduction} 
In this letter we investigate the possibility of neutrino gravitational  
lensing by astrophysical objects (stars, galaxies, or rather galactic halos).  
Unlike photons, neutrinos can cross even a stellar core.  
This results in a much better focalization due to 
the improvement in the lens quality.  
However, because of the extreme difficulty of detecting neutrinos and the poor  
angular resolution of ``neutrino telescopes'', one can only expect signal  
intensification rather than a spectacular lensing pattern like those of photons. 
The possibility of neutrino focusing by the Sun's core was previously  
debated in Ref.~\cite{Gerver:1988zs}. 
Neutrino lensing has also been used as a possible explanation for  
the time difference between neutrinos from supernova  
SN1987a \cite{supernova}. 
Here, we pursue an exhaustive and general analysis of neutrino lensing  
involving calculation of the neutrino deflection with different 
density profiles, a comprehensive discussion of neutrino absorption  
and signal magnification, and a description of possible  
applications. 
The letter is organized as follows. 
The gravitational deflection of neutrinos is calculated in  
Section \ref{rafel}.  
In Sect.~\ref{vero}, we study the neutrino interactions in a stellar medium  
and the depletion of the neutrino flux after crossing a massive object. 
Signal enhancement is estimated in Sect.~\ref{tools}. 
Finally, in Sect.~\ref{dominique}, we present some practical examples of our 
analysis. 
It is clearly shown that the Earth-Sun distance is too small for a sizable  
lensing effect to take place, although an observatory on Uranus would notice  
the enhancement of distant neutrino sources whenever they are aligned with  
the Sun. The case of galaxies, or rather their halos, is also 
contemplated. We finally consider binary systems, which are the 
most promising case. 
A more extensive and detailed description of our investigation, including  
complete calculations of the gravitational lensing effect, 
neutrino absorption and signal enhancement  
(taking into account the geometry, the amplification factor, and certain  
approximations of the lensing phenomena) can be found 
in Ref.~\cite{Escribano:1999gy}. 
 
\section{Gravitational deflection of neutrinos} 
\label{rafel} 
In this section we study the deflection of neutrinos from straight-line motion 
as they pass through a gravitational field produced by a compact object of  
mass $M$ and physical radius $R$.  
We will distinguish two cases:  
when the neutrino flux passes far away from the object (OUTside solution),  
a situation equivalent to the gravitational lensing of photons,  
and when the neutrino flux passes through the object (INside solution).  
In the latter case, we consider three specific cases depending on the compact 
object density profile:  
constant density (a profile fit for planetary objects),  
Gaussian density distribution (suitable for stars\footnote{ 
The density profile of stars is not exactly Gaussian but we use it here  
in order to obtain simple analytical results.  
Such a description should be considered as a good approximation to the  
real case.}) 
and Lorentzian density distribution (which could be associated with a galactic  
halo\footnote{The Lorentzian profile behaves as $1/r^2$ for large $r$, in  
agreement with velocity dispersion curves for galaxies and clusters.}). 
 
For the OUTside solution, calculating the trajectory of a massless  
neutrino (or with a mass very small compared to its energy) in the  
Schwarzschild  metric under the assumption that $M/r$ is everywhere small  
along the trajectory \cite{Schutz,Weinberg} gives as a result for the 
net deflection angle $\Delta\phi_{\rm OUT}=4M/b$. 
For the case of a neutrino flux passing through the object, one must first, in  
order to study the neutrino trajectory, look for the form of the space-time  
in the region inside the object.  
Here we restrict ourselves to the case of static spherically symmetric  
space-times and static perfect fluids. 
In the region inside the object, exact solutions to the relativistic equations  
are very hard to solve analytically for a given equation of  
state \cite{Schutz}. 
One interesting exact solution is the Schwarzschild constant-density inside  
solution, which we use here as a typical density profile for planetary  
objects. 
The net deflection as a function of the impact parameter $b$ is then 
\begin{equation} 
\label{phiINnetconstrhomaintext} 
\Delta\phi= 
\left\{ 
\begin{array}{ll} 
\frac{4M}{b} & {\rm if}\ b\geq R\\[2ex] 
\frac{4M}{b}\left(1-\sqrt{1-\frac{b^2}{R^2}}\right)     
+2\arcsin\left[\frac{b}{R}\left(1-\frac{M}{R}\right)\right]\\[2ex] 
\ +\frac{3M}{R}\frac{b}{R}\sqrt{1-\frac{b^2}{R^2}} 
  -2\arcsin\left\{\frac{b}{R}\left[1-\frac{3M}{2R} 
           \left(1-\frac{b^2}{3R^2}\right)\right]\right\} & {\rm if}\ b<R 
\end{array} 
\right. 
\end{equation} 
where the outside solution is also included for completeness. 
\begin{figure}[t] 
\centerline{\includegraphics{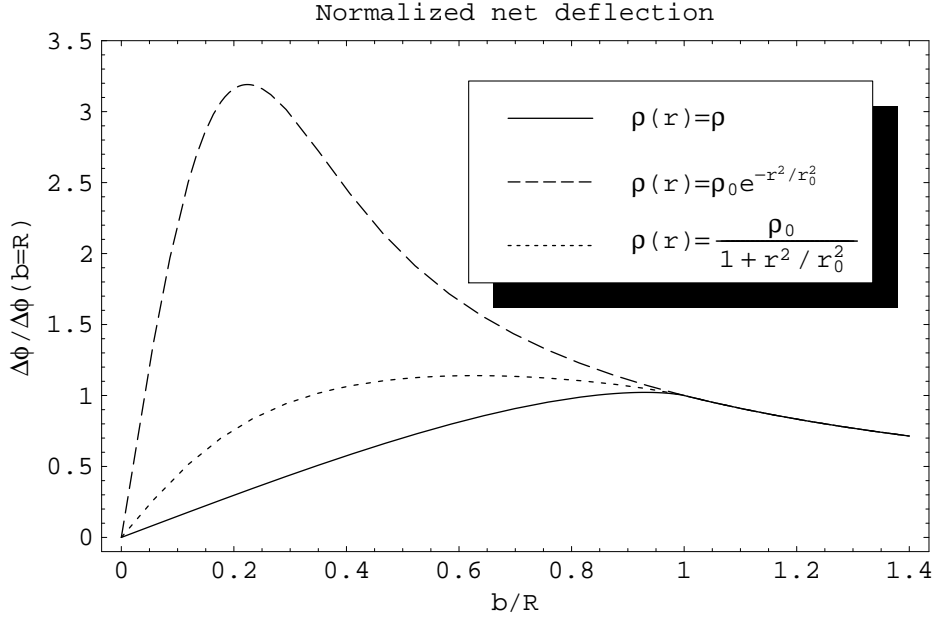}} 
\caption{Normalized net deflection $\Delta\phi/\Delta\phi(b=R)$ as a function  
of the normalized impact parameter $b/R$ for a constant density profile 
(solid line), a Gaussian density profile (dashed line) and a Lorentzian 
density profile (dotted line). In the last two cases $r_0$ is taken to be 
$r_0=0.2 R$.}  
\label{plotalldeflectnor}  
\end{figure}  
In Fig.~\ref{plotalldeflectnor}, we plot the normalized net deflection  
$\Delta\phi/\Delta\phi(b=R)$ as a function of the normalized impact parameter 
$b/R$ for the constant distribution density and compare it with other specific  
density profiles considered in the analysis. 
Such a normalization allows for a clear comparison of different profiles 
and is independent of the mass and physical radius of the compact object. 
 
We refer the reader to Ref.~\cite{Escribano:1999gy} for a detailed description 
of the calculations done in this section. 
 
Next, we analyze the solutions for the Gaussian and Lorentzian densities 
distribution. 
The Gaussian profile is a convenient approximation of the mass 
distribution in stars while the Lorentzian profile is valid for galactic  
halos. 
In both cases, it is possible to neglect the pressure with respect to the mass 
density, $p\ll\rho$ (see Ref.~\cite{Schutz} for the so-called Newtonian stars), 
and thus $4\pi r^3 p\ll m$. 
For the case of a Gaussian density profile $\rho(r)=\rho_0 e^{-r^2/r_0^2}$,  
the net deflection is 
\begin{equation} 
\label{phiINnetgaussianmaintext} 
\Delta\phi= 
\left\{ 
\begin{array}{ll} 
\frac{4M}{b} & {\rm if}\ b\geq R\\[2ex] 
\frac{4M}{b}\left(1-\sqrt{1-\frac{b^2}{R^2}}\right) 
+\frac{4M}{b}\frac{r_0/R\,e^{R^2/r_0^2}\sqrt\pi/2} 
{r_0/R\,e^{R^2/r_0^2}\sqrt\pi/2\,{\rm erf}\left(R/r_0\right)-1}\\[2ex] 
\times\left[\sqrt{1-\frac{b^2}{R^2}}\,{\rm erf}\left(R/r_0\right)- 
e^{-b^2/r_0^2}{\rm erf}\left(\sqrt{1-\frac{b^2}{R^2}}\frac{R}{r_0}\right) 
\right] & {\rm if}\ b<R 
\end{array} 
\right. 
\end{equation} 
where the error function is defined as 
${\rm erf}(z)=\frac{2}{\sqrt\pi}\int_0^z dt\,e^{-t^2}$. 
Fig.~\ref{plotalldeflectnor} shows the exact result in  
Eq.~(\ref{phiINnetgaussianmaintext})  
with the parameter $r_0$ fixed to $r_0=0.2 R$.  
As it is seen from Fig.~\ref{plotalldeflectnor},  
the maximal net deflection occurs at $b\simeq r_0$ and is  
$\Delta\phi(b\simeq r_0)\simeq 3.2\Delta\phi(b=R)$, 
showing that the lensing effect inside the star is bigger than the outside  
effect (except for $b\leq 0.04 R$).  
 
For the case of a Lorentzian density profile  
$\rho(r)=\frac{\rho_0}{1+r^2/r_0^2}$, the final deflection is 
\begin{equation} 
\label{phiINnetlorentzianmaintext} 
\Delta\phi= 
\left\{ 
\begin{array}{ll} 
\frac{4M}{b} & {\rm if}\ b\geq R\\[2ex] 
\frac{4M}{b}\left(1-\sqrt{1-\frac{b^2}{R^2}}\right) 
-\frac{4M}{b}\frac{1}{1-r_0/R\,\arctan(R/r_0)}\frac{r_0}{R}\\[2ex] 
\times\left[ 
\begin{array}{l} 
\sqrt{1-\frac{b^2}{R^2}}\arctan(R/r_0)\\[2ex] 
-\sqrt{1+\frac{b^2}{r0^2}} 
\arctan\left(\frac{\sqrt{1-\frac{b^2}{R^2}}}{\sqrt{1+\frac{b^2}{r0^2}}} 
             \frac{R}{r_0}\right) 
\end{array} 
\right] & {\rm if}\ b<R 
\end{array} 
\right. 
\end{equation} 
In Fig.~\ref{plotalldeflectnor}, we also plot the exact result in  
Eq.~(\ref{phiINnetlorentzianmaintext}) again with $r_0 = 0.2 R$. 
 
In summary, 
we conclude that only for the case of stars (Gaussian profile) is the lensing  
effect inside the star substantially amplified with respect  
to the outside effect  
(the inside effect should be compared with the effect at $b=R$). 
For galactic halos, the maximal inside net deflection is slightly bigger than 
at $b=R$, while for an object of constant density the inside lensing effect 
is always smaller than at $b=R$. 

\begin{figure}[t] 
\centerline{\includegraphics{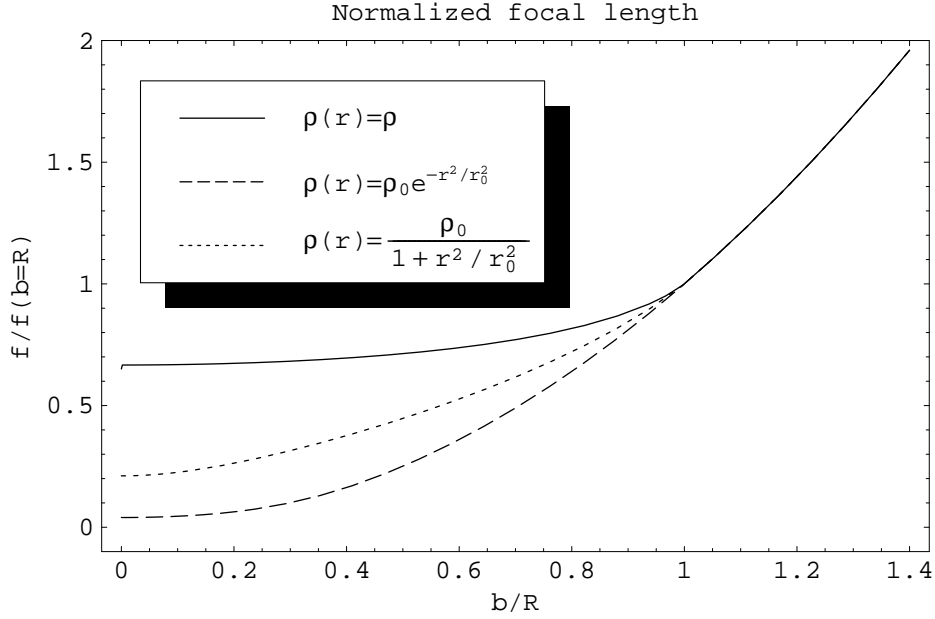}} 
\caption{Normalized focal length $f/f(b=R)$ as a function  
of the normalized impact parameter $b/R$.}  
\label{plotallfocallengthnor}  
\end{figure}  
Finally, we plot the focal length $f$ as a function of the 
impact parameter $b$.  
The effective focal length is defined as the distance at which the 
lens focuses the signal 
\begin{equation} 
\label{focallength} 
f(b)=\frac{b}{\Delta\phi}\ . 
\end{equation} 
A perfect lens would correspond to a constant focal length  
(independence of $b$). 
The neutrino flux would then be focalized in a single point with 
consequent signal intensification. 
In terms of the net deflection, a ``good lens'' requires that $\Delta\phi$ 
increases with the impact parameter $b$. 
In Fig.~\ref{plotallfocallengthnor}, the (normalized) focal lengths for the 
three different profiles are shown. 
We postpone to Sections.~\ref{tools} and \ref{dominique} our comments about
the quality of the different gravitational lenses considered here. 
  
\section{Neutrino absorption} 
\label{vero} 
As long as neutrinos are able to travel across a massive object,  
one has to consider their interactions with the matter inside. 
These interactions, dominated by scattering on  
nucleons \cite{Gandhi:1998ri}, will indeed reduce the neutrino flux and  
thus the efficiency of signal amplification due to lensing. 
Charged-current reactions convert the neutrinos into charged leptons 
that will later decay or be absorbed by matter while neutral-current 
reactions deviate them by large angles compared to the deflection angle. 
In both cases, the interacting neutrinos will not contribute to 
signal enhancement. 
 
The probability of transmission for a neutrino crossing at a distance $b$ 
from the center of the object  
is\footnote{We assume for simplicity an object with spherical symmetry.} 
\begin{equation} 
\label{probtrans} 
P_T(b,E_\nu)=\exp\left(-2\,\sigma^{\nu N}(E_\nu) 
             \int_b^R N_N(r)\frac{r\,dr}{\sqrt{r^2-b^2}}\right)\ , 
\end{equation} 
where $\sigma^{\nu N}(E_\nu)$ is the neutrino-nucleon cross section as a 
function of the laboratory neutrino energy and $N_N$ is the number density 
of scatterers for the case of nucleons. 
In order to compute the probability of transmission,  
one has to collect information not only on the scattering cross section  
but also on the mass density profile and the composition of the object  
(the latter is needed in order to relate number and mass densities). 
A detailed calculation of Eq.~(\ref{probtrans}) is found  
in Ref.~\cite{Escribano:1999gy}. 
\begin{figure}[t] 
\centerline{\includegraphics[width=0.85\textwidth]{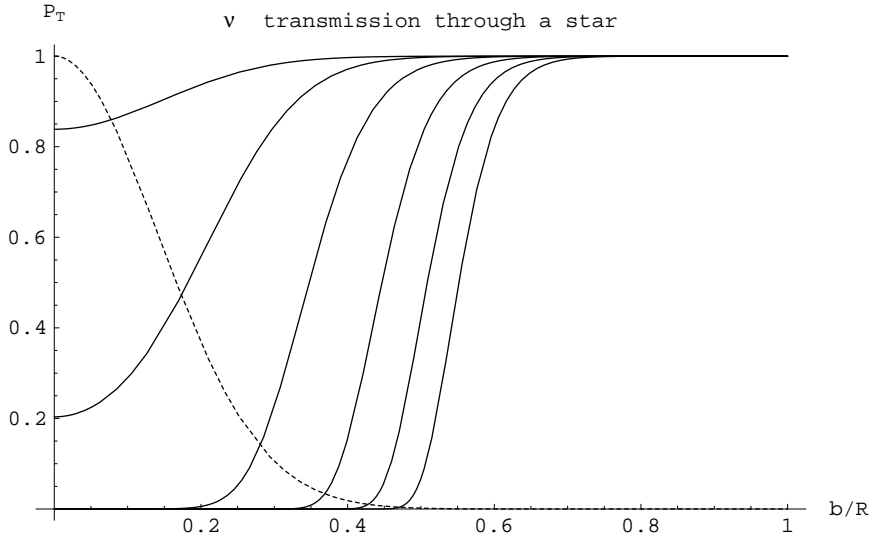}} 
\caption{Neutrino probability of transmission $P_T(b,E_\nu)$ as a function 
of $b/R$ for neutrinos crossing the Sun with energies  
$E_\nu=10$ GeV, 10$^2$ GeV, 10$^3$ GeV, 10$^4$ GeV, 10$^5$ GeV and 10$^6$ GeV  
(solid lines from left to right). 
The Gaussian density profile with $r_0=0.2 R_\odot$ is included for 
completeness (dashed line).}  
\label{plotabsorption}  
\end{figure}  

Let us now briefly discuss the results and consequences of  
Eq.~(\ref{probtrans}) for the two relevant physical cases studied  
in Sect.~\ref{rafel}, {\it i.e.}~neutrinos crossing a star 
or a galactic halo. 
For the case of a star, Eq.~(\ref{probtrans}) is 
calculated taking the Sun as example\footnote{Neutrino flux attenuation  
in the Sun has also been considered in Ref.~\protect\cite{Ingelman:1996mj}.}. 
The probability of transmission of neutrinos\footnote{A similar figure is 
obtained for antineutrinos. In this case, however, there also exists a 
contribution from $\bar\nu_e e$ scattering that dominates over the 
$\bar\nu N$ contribution at the Glashow resonance,  
$E^{\rm res}_{\bar\nu}=M_W^2/2m_e\approx 6.3\times 10^6$ GeV,  
but it is negligible at all other energies \cite{Gandhi:1998ri}.} 
across the Sun is then shown in Fig.~\ref{plotabsorption} as a function  
of $b/R$ for different incoming neutrino energies. 
From Fig.~\ref{plotabsorption}, it comes out that 
$E_\nu \geq$ 1 TeV neutrinos are completely absorbed or scattered away 
in the core of the star. Even at 100 GeV, matter interactions are 
frequent enough to significantly attenuate the flux leaving the star. 
Consequently, the region inside the star where lensing is more  
efficient (it was shown in Sect.~\ref{rafel} that the star's core acts 
as a ``good lens'') is ruled out as a lens for neutrinos with  
$E_\nu \geq$ 100 GeV due to flux attenuation. 
For $E_\nu <$ 100 GeV neutrinos, however, matter interactions have  
practically no incidence on the outgoing flux which is then recovered after 
lensing by the star. 
 
Neutrinos passing through a galaxy may interact either with its visible  
matter or with the surrounding halo of massive relic neutrinos. 
For the former, a rough estimate of the average density of stars in a  
galaxy gives 1 pc$^{-3}$, which results in a negligible probability 
for a neutrino to encounter a star during its passage through the galaxy, 
even if it traverses the whole disk and the bulge. 
For the latter, only interactions with ultrahigh energy neutrinos 
are significant \cite{Roulet:1993pz} and thus are not taken into account 
in our present framework. 
We conclude that the passage of neutrinos through a galaxy  
will not decrease their flux and hence not influence the lensing  
effect at all. 
 
\section{Signal enhancement} 
\label{tools} 
The difficulty of detecting neutrinos together with the poor angular  
resolution of ``neutrino telescopes'' leaves signal enhancement 
as the most likely (if not only) signature for neutrino lensing. 
Here we discuss in some detail this signal enhancement for different 
neutrino sources and lenses. 
 
\begin{figure}[t] 
\centerline{\includegraphics[width=0.85\textwidth]{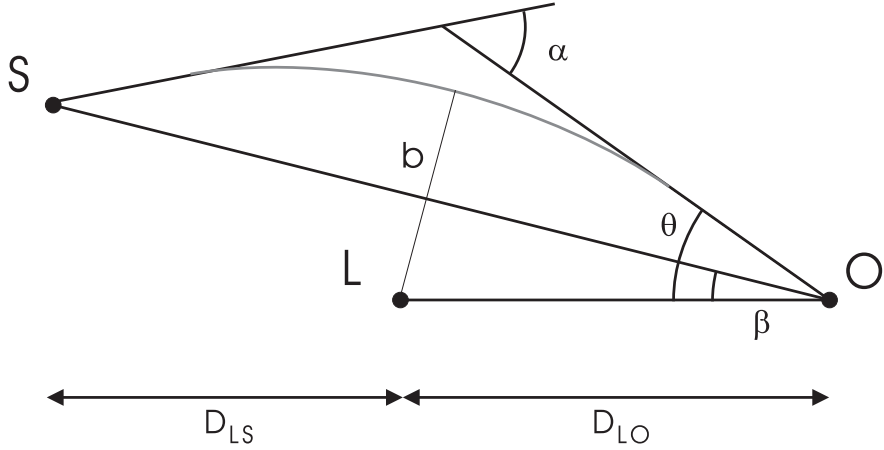}} 
\caption{The geometry of a lensing event. 
The lens is located at a distance $D_{\rm lo}$ from the observer and 
the source at a distance $D_{\rm so}$ (the distance between the lens 
and the source is $D_{\rm ls}$). 
The angular separation between the lens and the source is $\beta$ and 
the position of the image is at an angle $\theta$. 
The deflection angle is $\alpha$ (notice that we use $\Delta\phi$ for 
the deflection angle along the main text).} 
\label{figgeometry}  
\end{figure}

The total amplification (or magnification) of the signal intensity $\mu$  
for a point source is given by \cite{refgeo,Roulet:1997ur} 
\begin{equation} 
\label{ampli} 
\mu=\sum_i\frac{{\cal A}_i}{{\cal A}_0}=\sum_i 
\left|\frac{\theta_i d\theta_i}{\beta d\beta}\right|\ , 
\end{equation} 
where ${\cal A}_i$ and ${\cal A}_0$ are the surfaces of a certain image  
(in general there is more than one)  
and the source (both referred to the lens plane), $\theta_i$ are the angles  
of the source images with respect to the observer-lens line of sight,  
and $\beta$ is the angle to the actual source position. 
For a specific image with angle $\theta$, 
\begin{equation} 
\beta=\theta-\frac{D_{\rm ls}}{D_{\rm so}}\Delta\phi\ , 
\label{betadef} 
\end{equation}  
$\Delta\phi$ being the deflection angle and $D_{\rm ls}(D_{\rm so})$  
the distance between the source and the lens (observer) 
(see Fig.~\ref{figgeometry}). 

To observe neutrino lensing would require a huge amplification. 
As deduced from Eq.~(\ref{ampli}), this occurs only when the source, 
the lens, and the observer are in perfect alignment, $\beta=0$, or 
nearly aligned, $\beta\ll\theta_{\rm s}$, where $\theta_{\rm s}$ is  
the angular size of the source. 
When alignment occurs, the finite size of the source cannot be neglected 
and the amplification is then calculated as \cite{refgeo} 
\begin{equation} 
\label{amplisource} 
\mu_{\rm s}=\frac{1}{\pi\theta_{\rm s}^2} 
\int^{\theta_{\rm s}}_0 \mu(\beta) 2\pi\beta d\beta\ . 
\end{equation} 
In Eq.~(\ref{amplisource}) the detector is taken as pointlike. 
To consider its finite size\footnote{The case of a finite sized detector  
is described in Ref.~\protect\cite{Escribano:1999gy}.} would be relevant
only for $R_{\rm d}\gg R_{\rm s}\frac{D_{\rm lo}}{D_{\rm ls}}$, 
where $R_{\rm d}(R_{\rm s})$ is the size of the detector(source), 
that is for the case of very small sources or very nearby lenses. 
These finite size effects should be taken into account for the Sun 
in case it could act as a lens. 
 
Due to the complicated dependence of the net deflection $\Delta\phi$  
on the impact parameter $b$ (or, in terms of $\theta$, trough the relation  
$\theta=b/D_{\rm lo}$) for the different density profiles discussed in 
Sect.~\ref{rafel}, the amplification $\mu (\beta)$ as well as the  
function $\theta (\beta)$ are only known implicitly.  
The only exception is of course the OUTcase, where the deflection angle 
is simply $\Delta\phi=4M/b$. 
For this reason, we calculate here the signal magnification for two 
instructive approximations of $\Delta\phi$ in the INcase. 
The OUTcase is also shown for the sake of completeness. 
The results are the following: 
\begin{equation} 
\label{magniresults} 
\begin{array}{lll} 
\Delta\phi=\frac{4M}{b}\ , &  
\mu (\beta)=\frac{\theta_0}{\beta}\ , &  
\mu_{\rm s}=\frac{2\theta_0}{\theta_{\rm s}}\ ,\\[2ex] 
\Delta\phi=4M\delta\ , &  
\mu (\beta)=\frac{2\theta_0}{\beta}\ , &  
\mu_{\rm s}=\frac{4\theta_0}{\theta_{\rm s}}\ ,\\[2ex] 
\Delta\phi=4M\gamma b\ , &  
\mu (\beta)=\left(\frac{\theta_{\rm disk}}{\beta}\right)^2\ , &  
\mu_{\rm s}=\left(\frac{\theta_{\rm disk}}{\theta_{\rm s}}\right)^2\ ,\\[1ex] 
\end{array} 
\end{equation} 
where $\delta$ and $\gamma$ are constants with respect to $b$, $\theta_0$ is  
the value of $\theta$ for perfect alignment 
($\theta_0$ is nothing but the Einstein's radius, 
$R_{\rm E}\equiv 2\sqrt{\frac{M D_{\rm lo}D_{\rm ls}}{D_{\rm so}}}$, 
in angular units), and $\theta_{\rm disk}$ is the angle for which the 
linear approximation in $\Delta\phi$ holds. 
All the previous results are valid provided 
$\beta, \theta_{\rm s}\ll \theta_0, \theta_{\rm disk}$. 
The second column presents the magnification of a point source  
while the third one is for a finite sized source and for perfect alignment. 
The sum over multiple images is included. 
 
As stated in Eq.~(\ref{magniresults}), 
the magnification for a constant deflection is just twice that of the 
OUTcase.  
For a linearly increasing deflection the resulting magnification is  
instead the square of the OUTcase. 
The former case applies rather well to narrow Lorentzian 
profiles (small width, $r_0\ll R$, typical of galactic halos) 
while the latter fits correctly the central regions of the constant 
and Gaussian density profiles.  
This last case corresponds to the best amplification we can hope for  
since the signal crossing the central region is focused. 
This focalization is expected as soon as the focal length is almost  
constant in the central region, 
as seen in Fig.~\ref{plotallfocallengthnor}. 
Accordingly, the central region of  
planetary\footnote{The constant profile focuses a beam the most 
efficiently, as the focal length remains 
approximately constant over a wide range  of $b/R$.} 
and stellar objects acts as a ``good lens'' in the optical sense. 
Unfortunately, planetary objects are generally too small and light  
to be of practical use. 
Stars, however, can provide a huge signal amplification as well, as  
they also have a significant ``good lens'' region. 
We will now investigate in some detail the possible applications. 
 
\section{Applications} 
\label{dominique} 
As we are looking for huge amplifications, we consider the case of  
perfect alignment. 
$f(b)$ is the effective focal length for impact parameter $b$. 
Focalization on the observer will take place if $f(b)= 
f\equiv\left(\frac{1}{D_{\rm lo}}+\frac{1}{D_{\rm ls}}\right)^{-1}$. 
A good lens corresponds to a constant $f(b)$. 
 
\begin{description} 
\item[\emph{The Sun:}] 
We use as announced a Gaussian density profile as a reasonable  
approximation of matter distribution in the Sun. 
Unfortunately, when $f(b)$ is compared with  
$f\simeq D_{\rm lo}\equiv D_{\odot-\oplus}=1$ au 
(as $D_{\rm ls}\gg D_{\rm lo}$ is assumed), it appears that 
there is no intersection between the two curves. 
This means that the Sun cannot focus on the Earth a neutrino beam 
coming from a far source. 
The required intersection happens nevertheless at distances   
around 20 au, meaning that Uranians could perform neutrino lensing 
experiments using the Sun as a lens (see also Ref.~\cite{Gerver:1988zs}). 
For them, any neutrino source would be amplified in turn as the 
Sun sweeps in front of it! 
It is easy to check that Jupiter cannot replace the Sun as a 
useful lens for us, as its mass is about $10^{-3} M_\odot$. 
 
If one is not located at the focal point there is yet some 
amplification provided by the expression\footnote{We retain  
the notation $f(b)$ in Eq.~(\protect\ref{ampliinf}) so as not to be  
confused with $f$ although in this case $f(b)$ is a constant.} 
\begin{equation} 
\label{ampliinf} 
\mu_{\rm s}^{\rm Gauss}=\left(\frac{f(b)}{f(b)-D_{\rm lo}}\right)^2\ . 
\end{equation} 
Eq.~(\ref{ampliinf}) shows that in order to achieve an amplification 
factor greater than 2, $D_{\rm lo}$ must be in the range 
(0.3--1.7)$f(b)$. Outside this interval we can consider the 
amplification negligible. 
 
We conclude that the Sun is not a neutrino signal amplifier 
for experiments on Earth. 
 
\item[\emph{Stars, the lighthouse possibility:}] 
Stars are notorious lens candidates. 
Each alignment of a neutrino source with a star will provide a lensing 
event, {\it i.e.}~there exists an intersection between $f(b)$ and $f$. 
However, as $D_{\rm lo}$ is in this case at least of 1 pc 
(which is indeed the distance to the closest stars),  
$f(b)$ will always cross $f$ well outside the star\footnote{A simple 
calculation taking a Sun-like star as example and  
$D_{\rm lo}={\cal O}(1\,{\rm pc})$ gives  
$b=\sqrt{4M_\odot D_{\rm lo}}\simeq 20 R_\odot$ for the impact parameter.}.  
The case is thus similar to that of photons, namely the lens focuses 
a thin ring whose radius is the Einstein radius. 
According to Eq.~(\ref{magniresults}), a typical magnification factor for 
stars would be (assuming $D_{\rm ls}\gg D_{\rm lo}$) 
\begin{equation} 
\label{amplistars} 
\mu_{\rm s}^{\rm OUT}\approx 4\times 10^4\, 
\left(\frac{R_\odot}{R_{\rm s}}\right) 
\left(\frac{D_{\rm so}}{10\,{\rm kpc}}\right) 
\sqrt{\left(\frac{M}{M_\odot}\right) 
      \left(\frac{100\,{\rm pc}}{D_{\rm lo}}\right)}\ , 
\end{equation} 
where $R_{\rm s}$ is the physical radius of the source. 
Even if huge amplifications are possible, we have verified that for  
expected neutrino sources this signal enhancement is insufficient 
to allow the detection, the reason being that the magnification 
cannot compensate the $1/D_{\rm so}^2$ geometrical suppression of the  
signal. 
 
We shall now discuss the case of binary systems \cite{Roulet:1997ur}, 
where one star acts as a source and the other, very close to the 
first one, acts as a lens. 
Binary systems satisfy $D_{\rm ls}\ll D_{\rm lo}$ and 
$f\simeq D_{\rm ls}$. 
The most interesting behaviour occurs when the source is located  
at the effective focal length dictated by the central region of the  
lens; the image is then a full disk. 
This coincidence happens for 
\begin{equation} 
\label{coincidence} 
\frac{f(b)}{D_{\rm ls}}\approx 20\, 
\left(\frac{R}{R_\odot}\right)^2\left(\frac{M_\odot}{M}\right) 
\left(\frac{1\,{\rm au}}{D_{\rm ls}}\right)\approx 1\ , 
\end{equation} 
where $f(b)=r_0^2/4M$ in the central region of a Gaussian profile 
and the width is set at $r_0=0.2 R$. 
This condition can clearly be fulfilled. 
An upper estimate\footnote{The focal length is not exactly constant 
over the whole central region of the star but suffers from hyperbolic 
aberrations that indeed reduce the total  
magnification \protect\cite{Gerver:1988zs}.} of the total magnification is 
given by (see Ref.~\cite{Escribano:1999gy} for details) 
\begin{equation} 
\label{amplibinsys} 
\mu_{\rm  s}^{\rm Gauss}= 
\left(\frac{\theta_{\rm disk}}{\theta_{\rm s}}\right)^2= 
\left(\frac{r_0}{R_{\rm s}}\right)^2= 
\frac{1}{25}\left(\frac{R}{R_{\rm s}}\right)^2\ , 
\end{equation} 
where $\theta_{\rm disk}$ is the angular size of the disk image. 
The source should then be at least one order of magnitude smaller 
in radius than the lens in order to provide a large magnification. 
Binary stars will seldom meet all the above requisites 
but for more exotic systems, with a compact and intense neutrino source, 
this situation is really promising. 
In some way, the large companion acts as the lens focusing a lighthouse 
beam. 
 
\item[\emph{Galaxies:}] 
Lastly, we shall discuss another class of lens candidates, namely gala\-xies  
or rather galactic halos. 
In Fig.~\ref{plotalldeflectnor}, it is shown that the maximum deflection  
angle for a Lorentzian density profile is close to $4M/R$ and, more 
precisely, is always smaller than $\pi/2\cdot 4M/R$ (independently  
of the value taken for $r_0$). 
So, in order to focus on Earth, an effective focal length of 
\begin{equation} 
\label{fminLor} 
f(b)_{\rm min}\approx 30\,{\rm Gpc}\, 
\left(\frac{b}{R}\right)\left(\frac{R}{100\,{\rm kpc}}\right)^2 
\left(\frac{10^{12}M_\odot}{M}\right)\ , 
\end{equation} 
is needed. 
Therefore, since the radius of the Universe is about 5 Gpc,  
the neutrinos have to pass near the center of the galaxy,  
often through its visible part. 
Unlike photons, which are absorbed by dust clouds, 
neutrinos survive in their travel through the galaxy because the  
probability of meeting a star is tiny. 
For calculating the signal enhancement we use the Lorentzian density  
profile. In this case, however, the precise value of the width $r_0$ 
is not known, although it has to be small enough to explain the  
velocity dispersion curves inside galaxies. 
Assuming $r_0$ is small we calculate the magnification for $b>r_0$ 
(this time the image is a ring): 
\begin{equation} 
\label{ampligalaxy} 
\mu_{\rm s}^{\rm Lor}\approx 6\times 10^{-2}\, 
\left(\frac{100\,{\rm kpc}}{R_{\rm s}}\right) 
\left(\frac{100\,{\rm kpc}}{R}\right) 
\left(\frac{M}{10^{12}M_\odot}\right) 
\left(\frac{D_{\rm ls}}{1\,{\rm Gpc}}\right)\ . 
\end{equation} 
As in the case of binary systems, a small neutrino source is required 
for non-negligible signal enhancement. 
 
\item[\emph{Experimental prospects:}] 
Neutrino telescopes are putting limits on neutrino fluxes from  
point sources. 
Amanda quotes an upper limit on Earth of 
$\Phi_\nu\lsim\Phi_{\rm upper}=10^{-7}$ cm$^{-2}$ s$^{-1}$, 
assuming an $E^{-2}$ spectrum, for $E_\nu>10$ GeV and  
a declination larger than 30 degrees \cite{Andres:2000jf,Kim:1999kc}. 
Active galactic nuclei (AGNs), gamma-ray bursts (GRBs), 
supernova remnants (SNRs), emissions from accretion disks in binary systems,
etc.~are candidates for such sources 
(for a review, see Refs.~\cite{Gaisser:1995yf,Gandhi:1996tf}). 
Gravitational lensing increases the sensitivity to these sources but 
acts only fortuitously. 
 
A permanent amplification $\mu$ 
(we mean here that, relative to the source,  
the angular speed of the lens is negligible at our time scales) 
increases the expected number of events and therefore the sensitivity  
by the same factor $\mu$. 
However, to confirm that gravitational lensing is acting, 
the multiplicity of the image must be observed; 
it thus requires a photonic counterpart. 
 
Non-permanent lensing events will only be observed if enough  
interactions are tracked in the detector, 
{\it i.e.}~if  
$\int_{\rm lens.~ev.~}\!\!\mu(t)\,t\,dt\,\Phi_\nu\gsim
 T\,\Phi_{\rm upper}$, 
where $T$ is the total observation time and $\Phi _{\rm upper}$  
refers to the upper limit on point source flux reached by the experiment. 
The amplification pattern is in principle sufficient to demonstrate the  
gravitational lensing but requires enough statistics to establish the  
shape of the event. 
A coincidence with the photonic counterpart can again confirm the lensing  
effect. 
 
Finally, binary systems allow for periodic amplifications. 
The average sensitivity is increased by a factor 
$\int_{\rm rev.~}\!\!\mu(t)\,t\,dt$ and will in most cases improve only  
slightly (as the alignment time is always much shorter than a revolution). 
If the detector is not sensitive enough to match the unamplified source, 
it may collect the amplified signal periodically. 
It is worth pointing out that the neutrino signal is amplified while the  
source is hidden;  
the photon and neutrino fluxes have thus anticorrelated time dependences. 
It should be noted however that small sources are needed for efficient  
amplification. 
\end{description} 
 
\section{Conclusions} 
In this letter we have examined neutrino gravitational lensing by 
astrophysical objects. 
Unlike photons, neutrinos can cross a stellar core. 
We have calculated the deflection angle and the effective focal length 
for different density profiles. 
The neutrino absorption is also discussed and signal  
amplification is estimated for the relevant cases. 
Our formalism is then applied to stars and galaxies (or galactic halos). 
 
Our analysis shows that neutrinos passing through the  
central region of a star are deflected at an angle increasing 
linearly with the impact parameter. 
This results in a ``good lens'' that can focus neutrinos into a  
small region. 
The phenomenon operates up to large energies (about 300 GeV for the Sun) 
before the star becomes opaque to neutrinos. 
Unfortunately, focalization by the Sun occurs only at the distance of 
Uranus, so that amplification on Earth is modest. 
A more promising case is that of binary systems,  
where a large object would focus and ``beam'' neutrinos originating 
from a smaller source. 
 
\section*{Acknowledgements} 
Work partly supported by IISN Belgium and by the Communaut\'e  
Fran\c caise de Belgique (Direction de la Recherche Scientifique programme  
ARC). 
V.~Van Elewyck and D.~Monderen benefit from a FRIA grant. 
%Work also partly supported by the EEC, TMR-CT98-0169, EURODAPHNE network. 

\end{document}